\documentclass{article}

\usepackage{psfrag}

\usepackage{latexsym}
\usepackage{amsmath}
\usepackage{graphicx}

\makeindex

\graphicspath{{figure/}}

\date{ }

\begin{document}

\title{Hysteresis Models of Dynamic Mode Atomic Force Microscopes:
	Analysis and Identification}

\author{Michele Basso, Donatello Materassi\\
% }
% \affiliation{
	\small Dipartimento di Sistemi e Informatica, Universit\`{a} di Firenze\\
	 \small via S. Marta, 3, I-50139 Firenze (Italy)\\ 
	\small fax: +39-0554796363 \quad tel: +39-0554796524 \\ 
	(\small \texttt{basso@dsi.unifi.it / materassi@dsi.unifi.it})\\
% 	}
% }
% \author{
~\\
Murti Salapaka\\
% }
% \affiliation{
	\small Electrical Engineering Department, Iowa State University\\
	 \small 3128, Coover Hall, Ames, Iowa - 50011\\ 
	(\small \texttt{murti@iastate.edu})
% }
}

\maketitle


\begin{abstract}
% This article is a shame
A new class of models based on hysteresis functions is developed to describe 
atomic force microscopes operating in dynamic mode.
Such models are able to account for dissipative phenomena in the tip-sample 
interaction which are peculiar of this operation mode. 
% The models are capable of explaining intricate features of experimental data
% including dissipative aspects of the tip sample interaction.
The model analysis, which can be pursued
using frequency domain techniques, provides a clear insight of  specific nonlinear
behaviours. Experiments show good agreement with the identified models.
\end{abstract}

\section{Introduction}
\begin{figure}[hbt]
   \begin{center}
     \includegraphics[width=0.6\columnwidth]{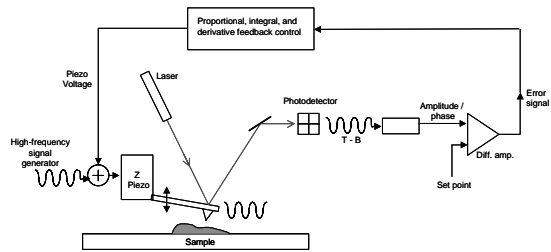}\\
     \caption{Schematic of a tapping-mode AFM. \label{AFMscheme}}
    \end{center}
\end{figure}

Physical systems showing impact phenomena are frequent
in many fields \cite{Brogliato}. Main applications occur 
in mechanics where macroscopic objects are considered. In
this case, an impulsive approximation for interaction forces
with a pure repulsive nature can often be correctly assumed.
Moreover, energy losses are traditionally considered by introducing
the concept of coefficient of restitution \cite{UrtiBrach} \cite{Fontaine}. 
However, there are many situations when this kind of approximation
can not be considered satisfactory, for example when the interaction
involves both attractive and repulsive
forces or when the interaction can not be assumed instantaneous.
The aim of this work is to exploit a hysteresis function
to model the related interaction forces. This new model 
can be viewed as a generalization of the impulsive case and allows
for the use of potential functions even if the system is dissipative. 
It also presents advantages when the interaction forces involve both
repulsive and attractive parts or when the duration of the impact is not
neglegible. In addition the hysteresis model allows for the use of 
powerful analysis techniques, such as harmonic balance \cite{Khalil}, 
which could not be used for impulsive forces.
To show how the above impact model can be successfully
employed, its application to an Atomic Force Microscope (AFM) is 
demonstrated.
Specifically, we limit ourselves to the study of an AFM operating in dynamic mode, whose schematic is
depicted in Figure \ref{AFMscheme}: the cantilever is periodically forced by a piezo
placed under its support inducing a periodic
oscillation that is influenced by the interaction forces
between the cantilever tip and the sample.
The topography can be inferred by slowly moving the cantilever
along the sample surface by means of a piezoactuator and by measuring the
amplitude of the cantilever deflection through an optical lever method.
A feedback controller driving the piezo input voltage is employed 
to reject variations of the separation between the sample and the tip center of oscillation
due to the sample topography.
In AFM, the amplitude-distance curve is not used to obtain the sample topography.
It is the control-signal that is used to get the image.
Therefore, the amplitude-distance curve for topography is not crucial.
However, topography is not the only information one might be interested in.
One of the primary uses of AFM is the study of force interactions \cite{israelachvili}. 
Two methods are prevalent. The cantilever-sample offset (also termed as separation), which is a measure of the distance of the cantilever holder and the piezo-positioner, 
is first reduced by the piezo positioner, where the sample surfaces approaches the cantilever-tip. In the retract phase the cantilever-sample offset is
increased by using the piezo positioner.  During the approach and the retract phases the cantilever deflection
signal is recorded. The force felt by the cantilever, can be obtained by multiplying the deflection by its
spring-constant. By plotting the force felt by the cantilever  against the cantilever-sample offset the force
curves are obtained. These curves are called static force curves. 
In dynamic force curves the cantilever is
oscillated using the dither piezo. The amplitude of the first harmonic is plotted against the cantilever-sample
offset  during the approach and retract phases. The dynamic force curves are gentler on the sample and
therefore are the preferred means of investigating samples that are soft (e. g. biological samples). One of the difficulties of using the
dynamic force curve mode when compared to static force curve mode in obtaining force
curves is that determining force-separation curves from the measured amplitude-separation
curves is not as straightforward.
In most cases, dynamic force curves are obtained by intensive numerical simulation.
For example, in \cite{Attrattive} and \cite{Salapaka2} models that accurately
describe the device behaviour are proposed.
In another approach, parametrized models of the tip sample interaction are assumed,
the parameters identified using the amplitude-separation data, and subsequently,
the force-curve data is generated using the identified model.
% <NEWS>
In \cite{Holscher2006} an identification algorithm
of the force-curve is obtained by the numerical computation
of an explicit integral equation.
% </NEWS>
There are only few attempts in the literature on analytical results.
Analytical results can be found in \cite{Salapaka1}, where a simple 
impulsive impact model is developed. However, since the employed model neglects
attractive forces, it does not seem able to explain some important
characteristics of the tip-sample interaction observed in experiments.\\
In this work, we develop a complete frequency analysis of a 
dynamic-mode AFM exploiting the proposed hysteresis description 
and taking into account attractive forces in the sample-cantilever 
interaction.
The main feature of the proposed model is to provide results
without the means of numerical simulations, for example evaluating the 
separation-amplitude curve for a large class of interaction forces comprising 
some of the common potential functions studied in the literature, 
such as the classical Lennard-Jones potential \cite{Cappella}.
Other peculiar attractive features of the proposed class of models are:
{\sl i)} it can easily account for energy losses;
{\sl ii)} it is suited for nonlinear frequency-domain identification techniques such as those proposed in \cite{Salapaka2} and \cite{coc97};
{\sl iii)} it facilitates to study some structural properties of the system such as
bifurcation phenomena experimentally observed exploiting
frequency domain techniques as in \cite{PredictBifurcations}.
Identification results based on experimental data are provided 
where the hysteresis model gives a good  qualitative and quantitative characterization 
of the tip-sample behaviour.\\
The paper is organized as follows.
In Section \ref{ModelCollision}  we briefly describe the general
problem of modelling an impact reminding many consolidate notions
for the sake of clarity.
In Section \ref{sect_AFMmodel}  we exploit such a model to describe the AFM tapping-mode dynamics 
and in Section \ref{sect_Harmonic_Balance}  a frequency analysis is provided using
harmonic balance techniques.
In Section \ref{sect_identification} the identification procedure is described and finally in
Section \ref{sect_expresult} experimental results are discussed.

\section{Hysteresis functions to model a collision}\label{ModelCollision}
%The way we are modelling an impact is extremely easy.\\
Let $P_1$ and $P_2$ be two material objects with masses $m_1$ and $m_2$,
respectively, moving along the $x$ axis, with position $x_1$ and $x_2$ 
($x_1 < x_2$). We consider $P_1$ and $P_2$ subject to external forces
$f_1(t)$ and $f_2(t)$ respectively, and to a mutual internal
force. 
% Let $h_1$ be the intensity of the internal force on $m_1$
% and $h_2$ the intensity on $m_2$, with $h_1 = -h_2$. We assume
% $h_1$ and $h_2$ depending only on time $t$, the
% relative position $x_2 - x_1$ and the relative velocity $\dot x_2 - \dot x_1$.
$P_1$ exerts a force on $P_2$ given by $h_2$ and $P_2$ exerts an equal and opposite force $h_1$ on  $P_1.$  The interaction force $h_i$ are dependent on time $t,$ relative separation $x_1-x_2$ and relative velocities $\dot{x}_1-\dot{x}_2$.
The following dynamical relations describe the system
\begin{equation} \label{Sistema_Generale}
  \left\{
    \begin{aligned}
      m_1 {\ddot {x}}_1=f_1(t)+h_1(t,x_2-x_1,\dot x_2-\dot x_1)  \\
      m_2 {\ddot {x}}_2=f_2(t)+h_2(t,x_2-x_1,\dot x_2-\dot x_1).
    \end{aligned}
  \right.
\end{equation}
Earlier interaction models usually neglected dissipation losses or used a constant coefficient of restitution to account for such losses.
Defining $\delta := x_2-x_1$, we suppose the interaction between the two masses
is negligible outside a time interval $[t_s, t_f ]$ where $\delta(t_s) =\delta(t_f)$.
We intend to limit our study to the case where $P_1$ and $P_2$ get closer at the
beginning, and then further. We assume that the system dynamics can be split in two different
phases: an ``approach phase'' in the time interval $[t_s, \bar t]$ where $\dot \delta \leq 0$ and a
``retract phase'' in the time interval $[\bar t, t_f]$ where $\dot \delta \geq 0$. We also consider
that $\delta(t)$ is a continuous function and that the set of points where $\dot \delta = 0$ has zero measure.
The model we intend to employ in this work defines a particular form for the interaction forces and, at the same time,
allows one to generalize the case of constant coefficient of restitution, not only for an instantaneous impact
time. In addition, it presents advantages in the study of impacting systems with
periodic behaviours. 
% The simple idea is to consider two different positional forces during the impact, the first
% one when the masses are approaching, the second one when the masses are departing 
% We account for energy
% losses by assuming a force $h$ that depends only on the sign of the relative velocity of the impacting bodies,
% that is
The interaction force assumes two different forms during the approach and the retract phases
\begin{equation}
  h(\delta,\dot \delta):= \left\{
                   \begin{array}{ll}
                        h^+(\delta) & \textrm{if {}} \dot \delta > 0 \\
                        h^-(\delta) & \textrm{if {}} \dot \delta < 0. \\
                        % \frac{h^+(\delta)+h^-(\delta)}{2} & \textrm{if {}} \dot \delta = 0. \\
                   \end{array}
                \right.
\end{equation}
The simplicity of the dependence of $h$ on the sign of $\dot \delta$ leads to a tractable
analysis while capturing the prominent features of finite time that can have both
attractive as well as repulsive forces.\\
If $h^+$ and $h^-$ are integrable, then the potential functions $U^+$ and $U^-$ can be introduced with
\begin{equation}
                   \begin{array}{l}
                        U^+(\delta) = -\int_{\delta(t_s)}^\delta h^+(x) \mathrm{d}x\\ % & \textrm{if {}} \dot d > 0 \\
                        U^-(\delta) = -\int_{\delta(t_s)}^\delta h^-(x) \mathrm{d}x.\\ % & \textrm{if {}} \dot d < 0
                   \end{array}
\end{equation}
As $h^-(\delta) > h^+(\delta) ~~\forall~  \delta < \delta(t_s)$, we have also that
$U^-(\delta) > U^+(\delta)~~\forall~\delta < \delta(t_s)$.
Considering the instant $\bar t$ when the relative distance $\delta(t)$ is the smallest, we can state that the
potential interaction energy is $U^-(t)$ if $t < \bar t$, while it is $U^+(t)$ if $t > \bar t$. 
At any instant the interaction force is conservative except at $t = \bar t$ when the relative velocity is
zero and we have an ``instantaneous'' energy variation $\Delta E$ equal to
\begin{equation} \label{Perdita_Isteresi}
  \Delta E = U^-(\delta(\bar t))-U^+(\delta(\bar t)).
\end{equation}
The energy lost in the impact can be interpreted as the area between the curves
$h^-(\delta)$ and $h^+(\delta)$ in the interval $[\delta(\bar t), \delta(t_s)]$. 
In the next section we will show how this class of hysteresis functions can be exploited
to analyze and identify tip-sample interactions in AFMs.

\section{AFM model}\label{sect_AFMmodel}
\begin{figure}[hbt]
	%\psfrag{g}[][l]{$\gamma(t)=\Gamma\cos(\omega t)$}
	%\psfrag{y}{$y$}
	%\psfrag{x}{ }
	%\psfrag{u}{ }
	%\psfrag{L}{$\mathcal{L}$}
	%\psfrag{xi}{ }
	%\psfrag{z}{ }
	%\psfrag{N}{$h(\cdot)$}
   \begin{center}
\setlength{\unitlength}{0.5mm}
\begin{picture}(160,80)
\thinlines
\put(32.5,50){\circle{5}}
%\put(25.25,55){\makebox(0,0){$+$}}
\put(32.5,50){\makebox(0,0){$+$}}
%\put(26.5,42.5){\makebox(0,0){$+$}}
\thinlines
%\put(15,50){\vector(1,0){15}}
\put(15,50){\vector(1,0){15}}
\put(35,50){\vector(1,0){25}}
\put(100,50){\vector(1,0){45}}
\put(127.5,50){\vector(0,-1){26.5}}
\put(125,21){\vector(-1,0){25}}
\put(60,21){\line(-1,0){27.5}}
\put(32.5,21){\vector(0,1){26.5}}
\put(127.5,21){\circle{5}}
\put(127.5,21){\makebox(0,0){$+$}}
\put(145,21){\vector(-1,0){15}}
\put(148,22){\makebox{{ $l$}}}
\put(105,23){\makebox{{ $\delta(t)$}}}
\put(60,11){\framebox(40,20){{$h(\cdot)$}}}
\put(60,40){\framebox(40,20)
{{$\cal L$}}}
\put(148,50){\makebox{{ $y(t)$}}}
\put(-5,50){\makebox{{ $\gamma(t)$}}}
%\put(80,4){\makebox(0,0){nonlinear subsystem}}
%\put(80,68){\makebox(0,0){linear subsystem}}
\end{picture}
     \caption{A feedback interconnection of a linear system and a nonlinear static function.
			 \label{LureSystem}}
    \end{center}
\end{figure}
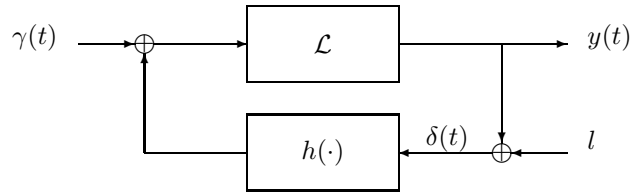

AFM cantilevers can be modeled as a feedback interconnection of a linear system $\mathcal{L}$ and a nonlinear static function $h$
as depicted in Figure \ref{LureSystem}.
Models with this peculiar structure are well-known as Lur'e models \cite{Khalil}.
The system equation can be conveniently written using the symbolic form
\begin{equation}\label{LureEquation}
	y(t)=L\left(\frac{d}{dt}\right) [h(\delta(t),\dot\delta(t))+\gamma(t)]
\end{equation}
where $y(t)$ is the measured output (that is the cantilever tip deflection), $l$, apart an additive constant,
is the separation, $\delta(t)=y(t)+l$ represents the tip-sample distance, and $\gamma(t)$ 
is the external periodic forcing
$$
\gamma(t)=\Gamma\cos(\omega t+\phi)~~.
$$
% From a physical point of view,
%perspective,
The subsystem $\mathcal{L}$ describes the free cantilever dynamics, 
whose frequency response $L(i\omega)$ can be
precisely identified using thermal noise or a simple frequency sweep excitation
when the sample is absent \cite{Thermal}.\\
The feedback subsystem $h$ accounts for the sample interaction force, which is a highly nonlinear function of the tip-sample distance $\delta$. Modeling $h$ is still a challenging
task. The main difficulty lies in the choice of a suitable class of functions to describe the force potential.
% <NEWS>
It is a common choice to consider $h$ as the sum of a
conservative force $h_{con}$ and a dissipative one 
$h_{dis}$
\begin{equation}
	h(\delta,\dot \delta)=h_{con}(\delta) +
		h_{dis}(\delta,\dot \delta)
\end{equation}
giving to $h_{dis}$ a simple form to allow easy computation \cite{Lee2006}.
Also in \cite{Lee2006} it is proposed
\begin{equation}
	h_{dis}(\delta,\dot \delta)=\Gamma(\delta) \dot \delta
\end{equation}
where $\Gamma$ represents a sort of damping coefficient.\\
% </NEWS>
In this paper we consider the following class of hysteresis functions
which generalizes the one presented in \cite{Garcia2006}
\begin{equation}\label{generalclass}
   h(\delta,\dot \delta) = \left\{ 
 			\begin{array}{l}
				\sum\limits_{n=1}^{N} K_n^- h_n(\delta)\qquad 
					\mathrm{if}~\dot \delta < 0 \\
				\sum\limits_{n=1}^{N} K_n^+ h_n(\delta)\qquad 
					\mathrm{if}~\dot \delta \ge 0 \\
			\end{array}
		\right.
\end{equation}
where $h_n(\delta)$ are a class of suitable non-negative functions where the dependence 
on $\dot \delta$ occurs in $h$ considering only its sign as described in the previous section.
Relation (\ref{generalclass}) represents a vector space of hysteresis functions made of two 
different positional forces: the first one acts when the tip and the sample are approaching and the second one when they are getting further.
In order to make the interaction described by $h(\delta,\dot \delta)$ dissipative,
some constraints on the parameters $K^+_n$ and $K^-_n$ can be imposed.
For example, the condition 
\begin{equation}\label{constraints}
	K_n^-\geq K_n^+ 
\end{equation}
makes every base element $h_n$ dissipative. \\
In our analysis, we will consider two special cases of this hysteretic interaction.
% <NEWS>
This way of modeling dissipations has already been proposed
in \cite{MatBasGenCDC2004}
and has been successfully exploited by \cite{Garcia2006} in
an identification procedure.
% </NEWS>

\subsection{Piecewise interaction force \label{PWinteraction}}
\begin{figure} [ht]
	\psfrag{-Fa}{$K_1^-$}
	\psfrag{-Fb}{$K_1^+$}
	\psfrag{-Ka}{$K_1^--K_2^-(y+l)$}
	\psfrag{-Kb}{$K_1^+-K_2^+(y+l)$}
	\psfrag{y}{$y$}
	\psfrag{hydy}{$h$}
	\psfrag{-l}{$-l$}
   \begin{center}
      \includegraphics[width=0.4\columnwidth]{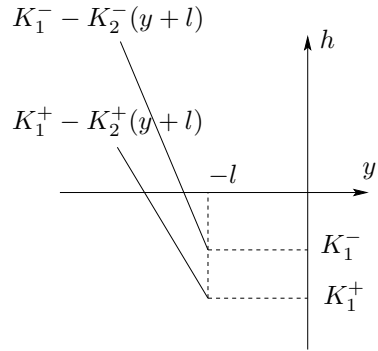}
   \end{center}
   \caption{Interaction force modeled by a piecewise linear function.
			\label{PWhysteresis}}
\end{figure}

The first class of potential functions we treat contains the functions $h(\cdot)$
in the form (\ref{generalclass}) where $N=2$ and
\begin{equation}\label{Isteresi_Lineare_Attrattiva_Repulsiva}
   h_n(\delta) = \left\{
     \begin{array}{ll}
       0          & \quad \mathrm{if}~\delta \geq 0 \\
       |\delta|^{n-1} & \quad \mathrm{if}~\delta < 0.  \\
     \end{array}
  \right.
\end{equation}
In Figure \ref{PWhysteresis} the shape of such a kind of functions is depicted.
Since $\delta=y+l$, here the parameter $l$ models the cantilever deflection at which the tip-sample 
interaction forces become effective.
% In this model, $l$ is an additional parameter that can be ea
% $K_1^-$ and $K_1^+$ take into account the attractive component of the interaction while $K_2^-$ and
% $K_2^+$ model the repulsive one.

\subsection{Lennard-Jones-like interaction force}
The Lennard-Jones potential
\begin{equation}
   h(\delta)=\frac{K_{n_a}}{\delta^{n_a}}+\frac{K_{n_r}}{\delta^{n_r}}\qquad 
		n_a<n_r \in \mathcal{N};~ K_{n_a},K_{n_r}\in \mathcal{R}
\end{equation}
is a common choice when fitting statically measured curves often used as a model of interaction 
potential between atoms (see \cite{israelachvili})
We consider a generalization of the Lennard-Jones Potential in the form (\ref{generalclass})
where
\begin{equation}\label{LJlike}
	h_n(y)=\frac{1}{\delta^n}.
\end{equation}
% These functions are a generalization of the static Lennard-Jones potential since they are 
% capable to consider energy dissipations by the means of a simple hysteresis law.
\begin{figure} [ht]
%	\psfrag{y}{$h(y,\dot y)$}
%	\psfrag{x}{$y$}
%	\psfrag{Distance}{\small Distance $\delta$}
	\psfrag{h}{$h$}
	\psfrag{y}{$y$}
	\psfrag{l}{$l$}
   \begin{center}
      \includegraphics[width=0.7\columnwidth]{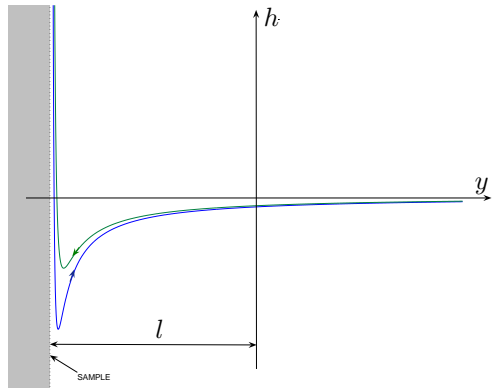}
   \end{center}
   \caption{Sketch of a Lennard-Jones-like interaction with hysteresis.\label{LJhysteresis}}
\end{figure}
\noindent The parameter $l$ represents the cantilever deflection  where the interaction force becomes infinitely large (Figure \ref{LJhysteresis}).
% <NEWS>
The choice of this class of functions is motivated by its semplicity
and also by the fact that long-range dissipative
interfacial forces has already been successfully modelled
using a time-dependent power law where the strength
of the force depends only on whether the probe approaches
or retracts away from the sample.
\cite{Garcia2006}
% </NEWS>

\section{Frequency Analysis via Harmonic Balance \label{sect_Harmonic_Balance}}
The linear part of the Lur'e system given by $\mathcal{L}$ in (\ref{LureEquation}) typically shows a sharp filtering effect beyond the first resonance peak because 
of a very high quality factor of the cantilever.
Indeed, it is experimentally observed that the cantilever trajectory has a quasi-sinusoidal behaviour. The cantilever-tip motion can be approximated by 
\begin{equation}\label{y_1(t)}
y(t) \simeq y_1(t) :=\mathrm{Re}[A+B \mathrm{e}^{i\omega t}] =A+B\cos(\omega t)
\end{equation}
The corresponding output of the nonlinear hysteresis block can be approximated as
\begin{equation}
h(y+l,\dot y) \simeq h(y_1+l,\dot y_1)\simeq 
	\mathrm{Re}\left[N_0 A + N_1 B  \mathrm{e}^{i\omega t} \right]
\end{equation}
where 
\begin{equation}
\label{general describing functions}
   \left\{
      \begin{aligned}
     & N_0=N_0(A,B,\omega):=\frac{1}{A}\frac{1}{T}\int_0^T h\big(y_1(t)+l,\dot y_1(t)\big) \mathrm{d}t\\
         & N_1=N_1(A,B,\omega):=\frac{1}{B}\frac{2}{T}\int_0^T h\big(y_1(t),\dot y_1(t)\big) \mathrm{e}^{-i\omega t} \mathrm{d}t
      \end{aligned}
   \right.
\end{equation}
are the constant and harmonic gains of
the nonlinear block also known as the describing functions of
the nonlinearity \cite{Khalil}.
We remark that $N_0 A$ and $N_1 B$ are the first two Fourier coefficients of $h(y_1(t)+l, \dot y_1(t)) $, thus
expression (\ref{general describing functions}) represents a first order harmonic truncation.
For the general class of hysteretic force models introduced, we obtain
\begin{equation} \label{N_0}
 \begin{aligned}%{ll}
   N_{0}&=\frac{1}{2\pi A} \sum_{n=1}^{N} 
   \left(   
   \int_{-\pi}^{0} K_n^+ h_n(l+ A+B\cos\tau) \mathrm{d}\tau+
   \int_{0}^{+\pi} K_n^- h_n(l+ A+B\cos\tau) \mathrm{d}\tau
		\right)=\\
     &=\frac{1}{2\pi A}\sum_{n=1}^{N}\Sigma_n \int_{0}^{+\pi} h_n[B(q+\cos\tau)] \mathrm{d}\tau \nonumber
 \end{aligned}
\end{equation} 
where 
\begin{equation}\label{Sigma}
\Sigma_n:=K_n^-+K_n^+
\end{equation}
and 
\begin{equation}
\label{qdefinition}
q:=\frac{l+A}{B}~.
\end{equation}
Similarly, we find for $N_1$
\begin{equation}\label{N_1}
 \begin{aligned} %{ll}
   N_{1}&=\frac{1}{\pi B} \sum_{n=1}^{N} 
   \left(   
   \int_{-\pi}^{0} K_n^+ h_n[B(q+\cos\tau)] \mathrm{e}^{-i\tau}\mathrm{d}\tau+
	 \int_{0}^{+\pi} K_n^- h_n[B(q+\cos\tau)] \mathrm{e}^{-i\tau}\mathrm{d}\tau
		\right)=\\
     &=\frac{1}{\pi B}\sum_{n=1}^{N}\left( 
     \Sigma_n \int_{0}^{+\pi} h_n[B(q+\cos\tau)]\cos\tau  \mathrm{d}\tau -
    i\Delta_n \int_{0}^{+\pi} h_n[B(q+\cos\tau)]\sin\tau  \mathrm{d}\tau	\right)
 \end{aligned}
\end{equation}
where 
\begin{equation}\label{Delta}
 \Delta_n:=K_n^+-K_n^-~.
\end{equation}
\normalsize
Substituting in (\ref{LureEquation}), assuming a sinusoidal forcing
$\gamma(t)=\mathrm{Re}[\Gamma\mathrm{e}^{i(\omega t + \phi)}]$, yields
% can be computed through the
% classical describing function method which originates the following
% equation  to be solved in $A$, $B$, $\phi$
\begin{equation}\label{Classical Autosostentamento}
   A+B\mathrm{e}^{i\omega t}= -L(0) N_0 A + L(i\omega) [-N_1 B + \Gamma \mathrm{e}^{i\phi}]\mathrm{e}^{i\omega t}\qquad\forall t
\end{equation}
or, equivalently,
\begin{equation}\label{autosostentamento} \left\{ \begin{array}{l}
            [1+L(0)N_0(A,B)]A=0\\
            \left[ 1+L(i\omega t)N_1(A,B) \right] B = L(i \omega
            t) \Gamma \mathrm{e}^{i \phi}. \\
        \end{array}
\right.
\end{equation}
Finally, we can easily decouple the variable $\phi$ from (\ref{autosostentamento}) as follows
\begin{equation}\label{autosostentamento2} \left\{ \begin{array}{l}
            [1+L(0)N_0(A,B)]A=0\\
            \left| 1+L(i\omega t)N_1(A,B) \right| B = |L(i \omega
            t)| \Gamma. \\
	    \phi=\arg\left[
			 L(i\omega)^{-1} +N_1(A,B)
		\right].
        \end{array}
\right.
\end{equation}
The equations (\ref{autosostentamento2}) represent a system of three nonlinear equations in the three
unknown $A, B, \phi$. By solving it, we can find the sinusoidal approximation of $y(t)$ given by (\ref{y_1(t)}).

\subsection{Piecewise interaction model analysis}
For the piecewise-linear potential described in  Section \ref{PWinteraction} we obtain
% \hspace{-0.1cm}
\begin{equation}
 \left\{
  \begin{array}{ll}
   N_0=&\frac{1}{A}\left[\Sigma_1 R_1(q)+\Sigma_2 R_2(q)B\right]\\
   N_1=&\frac{1}{B}\left[\Sigma_1 S_1(q)+i\Delta_1 T_1(q)\right]+\left[\Sigma_2 S_2(q)+i\Delta_2 T_2(q)\right]
  \end{array}
 \right.
\end{equation}
where
\begin{equation}
   \begin{aligned}
      R_1(q):=\frac{\mathrm{acos}(q)}{2\pi} &\qquad
      R_2(q):= \frac{q\mathrm{acos}(q)-\sqrt{1-q^2}}{2\pi}\\
      S_1(q):=-\frac{\sqrt{1-q^2}}{\pi} &\qquad
      S_2(q):=\frac{\mathrm{acos}(q)-q\sqrt{1-q^2}}{2\pi}\\
      T_1(q):=\frac{1-q}{\pi} &\qquad 
      T_2(q):=-\frac{(1-q)^2}{2\pi}
   \end{aligned} \nonumber
\end{equation} 
Finally, by the substitutions
\begin{equation}
  \begin{aligned}
     \chi(q):=\Sigma_1 R_1(q)  &\qquad  \Omega(q):=\Sigma_2 R_2(q)\\
     \Phi(q):=\Sigma_1 S_1(q)+i\Delta_1 T_1 (q) &\qquad
         \Psi(q):=\Sigma_2 S_2(q)+i\Delta_2 T_2(q)
  \end{aligned} \nonumber
\end{equation}
we can obtain for the describing functions
\begin{equation}
  \begin{aligned}
    N_0=&\frac{1}{A}[\chi(q)+\Omega(q)B]\\
    N_1=&\frac{\Phi(q)}{B}+\Psi(q).
  \end{aligned} \nonumber
\end{equation}
% Note that $N_0$ and $N_1$ are not dependent on the
% frequency $\omega$ and that $N_1$ can be formally expressed as a
% function of $q$ only. 
% In the sequel we will indicate it as $N_1(q)$.
In this model, the variable $q$ represents the ``penetration'' of the tip into the
sample. In fact, assuming as exact the first harmonic approximation,
we have that for $q>1$ the tip does not get in contact with the
sample; for $q=1$ the tip grazes the sample and for $q<1$ the tip enters
the sample. The case $q<-1$ does not have a physical meaning
in this model.
From (\ref{autosostentamento}), it is also possible to write $B$ as a function of $q$. 
In fact
\begin{equation}
    B = \frac{\Gamma}{\left| L(i\omega)^{-1} + N_1\right|}
\end{equation}
\normalsize
implies
\small
\begin{equation}
\left|L(i\omega)^{-1}B+B N_1 \right|^2=
   |L(i\omega)^{-1}B +\Phi +\Psi B|^2=\Gamma^2. \\
\end{equation}
\normalsize
The substitutions $\hat \Phi:=\Phi$ and
$\hat \Psi:=\Psi+L(i\omega)^{-1}$ yield
\begin{equation}
   \begin{array}{l}
      (\hat \Phi + \hat \Psi B)(\hat \Phi^* + \hat \Psi^* B)=\Gamma^2    \Rightarrow\\
      |\hat \Psi|^2B^2+2\mathrm{Re}[\hat\Phi \hat\Psi^*]B+|\hat \Phi|^2-\Gamma^2=0.
   \end{array}
\end{equation}
which is a simple second order algebraic equation whose roots are
\begin{equation}
	B
	(q)
	=
	%\Theta(q):=
		\frac{-\mathrm{Re}[\hat\Phi \hat\Psi^*]\pm
			\sqrt{\mathrm{Re}[\hat\Phi \hat\Psi^*]^2-|\hat \Psi|^2 (|\hat \Phi|^2-\Gamma^2)}}
		{  |\hat \Psi|^2  }.
\end{equation}
Substituting in (\ref{autosostentamento2}) and reminding that $l=qB-A$, we can finally write
% \begin{equation}\label{autosostentamento esplicito}
% 	\left\{\begin{array}{l}
% 		A(q)=-L(0)\left[\chi(q)+\Psi(q)\Theta(q)\right]\\
% 		B(q)=\Theta(q)\\
% 		\phi(q)=\arg\left[
% 					L(i\omega)^{-1} +N_1(A,\Theta(q))
% 					\right]\\
% 		l(q)=q\Theta(q)+L(0)\left[\chi(q)+\Psi(q)\Theta(q)\right]
% 		\end{array}
% 	\right.
% \end{equation}
\begin{equation}\label{autosostentamento esplicito}
	\left\{\begin{array}{l}
		A(q)=-L(0)\left[\chi(q)+\Psi(q)B(q)\right]\\
		% B(q)=\Theta(q)\\
		\phi(q)=\arg\left[
					L(i\omega)^{-1} +N_1(A,B(q))
					\right]\\
		l(q)=qB(q)+L(0)\left[\chi(q)+\Psi(q)B(q)\right].
		\end{array}
	\right.
\end{equation}

The variable $q$ (\ref{qdefinition}),
depends on $A$, $B$ and $l$ (see (\ref{qdefinition})), therefore equations (\ref{autosostentamento esplicito}) 
are implicit relations.
System (\ref{autosostentamento esplicito}) can not be solved in closed form
since it involves transcendental equations. However, it is possible
to obtain its solution through a conceptually easy method. Assuming that
$l$ is a known parameter of the model, it is possible by the last 
of (\ref{autosostentamento esplicito}), to determine the corresponding values 
of $q$ and then $A$, $B$ and $\phi$ by exploiting the remaining equations. 
In other words, we have transformed the problem of solving the whole system
(\ref{autosostentamento}) into the easier problem of solving a single
real equation in the unknown $q$.\\
%However, numerical methods are required.
Experimentally, the separation-amplitude curve is obtained by slowly
moving the sample towards the cantilever and measuring both the
amplitude of the first harmonic and the separation. Although it is
not possible to derive an explicit analytical form for $B=B(l)$, we
can give a parametric form for it. By using the ``$q$-explicit''
equations in (\ref{autosostentamento esplicito}) we can consider the
parametric curve
\begin{equation}
\label{(B,l) parametrica}
  \left\{
    \begin{aligned}
       & l=l(q)\\
       & B=B(q)
    \end{aligned}
  \right.
  \qquad \forall~ q \in \mathbf{R}.
\end{equation}

% In Figure \ref{fig_B_phi_l_attrattivo} we can observe the comparison between
% simulated results and the predicted ones using the analytical expression 
% (\ref{(B,l) parametrica}). The two curves show a good agreement.
% \begin{figure} [ht]
%    \begin{center}
%       \includegraphics[width=0.70\columnwidth]{Simulazione6}
%       %\includegraphics[width=0.40\columnwidth]{phi_l_attrattivo}
%    \end{center}
%    \caption{Comparison between the separation amplitude curve obtained analytically  and
% 		by a simulation.}
%    \label{fig_B_phi_l_attrattivo}
% \end{figure}
% In addition, we note that there is a range of values of $l$ for which three
% different periodic solutions exist, two stable and one unstable. This
% critical behaviour, exhibited for some values of the parameters, generates ``jumps''
% %(dotted lines in Fig. \ref{fig_B_phi_l_attrattivo})
% depending on the scan direction ($l$ increasing or decreasing).
% Both experiments and simulations confirm this sort of phenomena  \cite{Attrattive}.
% The proposed approach has allowed, in addition, to detect the unstable
% periodic solutions which are difficult to obtain by simulation. A
% similar behaviour can be observed in the phase-separation diagram and
% can be explained exactly in the same way.
% In addition, even in this simple model, we can observe that a relation of
% ``quasi-linearity'' exists between amplitude and separation \cite{Salapaka1}.
% and
%between $\mathrm{sin}(\phi)$ and separation.

\subsection{Lennard Jones-like hysteretic model analysis}
For the generic hysteretic interaction force of the class (\ref{LJlike}), we can evaluate 
the describing functions $N_0$ and $N_1$ of the nonlinearity $h$:
\begin{equation}
  \nonumber
   \left\{
     \begin{array}{ll}
   N_{0}&=\sum_{n=1}^{N}\frac{\Sigma_n}{A B^n} R_{n}(q)\\
   N_{1}&=\sum_{n=1}^{N}\frac{1}{B^{n+1}}\left[\Sigma_nS_{n}(q)+i \Delta_n T_n(q)\right]
    \end{array} 
    \right.
\end{equation}
where the functions
\begin{equation}
 \begin{aligned}
  &R_n(q):=\frac{1}{2\pi} \int_{0}^{\pi} \frac{1}{(q+\cos\tau)^n}\mathrm{d}\tau &\\
  &S_n(q):=\frac{1}{\pi} \int_{0}^{\pi} \frac{\cos\tau}{(q+\cos\tau)^n}\mathrm{d}\tau &\\
  &T_n(q):=\frac{1}{\pi} \int_{0}^{\pi} \frac{-\sin\tau}{(q+\cos\tau)^n}\mathrm{d}\tau. &
 \end{aligned}
 \nonumber
\end{equation}
can be analytically evaluated for any given $n$ and $q>1$.

Imposing harmonic balance, we get 
\begin{equation}\label{HarmonicBalance}
  \left\{
     \begin{array}{l}
        A = -L(0)\sum_{n=1}^{N} \Sigma_n \frac{R_n(q)}{B^n} \\
        \left[ \Gamma \mathrm{e}^{i\phi}- \sum_{n=1}^{N} 
            \frac{\Sigma_nS_n(q)+i\Delta_nT_n(q)}{B^{n}} \right] L(i\omega)=B.\\
     \end{array} 
  \right.
\end{equation}
The second equation of (\ref{HarmonicBalance}) can be expressed in the form
\begin{equation}
   L(i\omega) \Gamma \mathrm{e}^{i\phi}= L(i\omega) 
	\sum_{n=1}^{N} \frac{\Sigma_nS_n(q)+i\Delta_nT_n(q)}{B^{n}}
		+B.
\end{equation}
We can remove $\phi$ by multiplying each term by its conjugate.
Finally, multiplying by $B^{2N}$ the equation can be easily rewritten as a $(N+2)$-degree polynomial in the variable $B$ whose coefficients depend only on the variable $q$
\begin{align} \label{polynomial}
   p(B)=\sum_{n=1}^{2N+2} C_n(q) B^n=0.
\end{align}
It can be shown that $C_{2N+2}=1/|L(i\omega)|^2$, $C_{2N+1}=0$ and $C_{2N}=-\Gamma^2$.
For sufficiently large $q$ (that is when the interaction is negligible) we  have that $C_{k}\cong 0$, $\forall k<2N$, therefore
\begin{equation}
   p(B) \cong (|L(i\omega)|^{-2} B^2 -\Gamma^2)B^{2N} =0.
\end{equation} 
One root of the equation above is $B\cong\Gamma |L(i\omega)|$. This solution corresponds to the free
oscillation amplitude that the cantilever assumes when the sample is far away and does not influence the
cantilever dynamics.
For every $q>1$, the polynomial equation (\ref{polynomial}) can be solved in $B$. Only the solutions
that are real and positive have relevance.
The constant component of the periodic solution $A$ can be evaluated exploiting the first of (\ref{HarmonicBalance}). The phase $\phi$ can also be similarly obtained as a function of the parameter $q$
\begin{equation}
   \phi(q)= \mathrm{arg}\left\{ L^{-1}(i\omega)+\sum_{n=1}^{N} 
		\frac{\Sigma_nS_n(q)+i\Delta_nT_n(q)}{B^{n+1}} \right\}.
\end{equation}
Finally, the parameter $l$ is given by the original relation
\begin{equation}
   l(q)=qB(q)-A(q).
\end{equation}
The final result is that the variables $A$, $B$, $\phi$ and $l$ are all expressed with respect to the parameter $q$. The separation-amplitude diagram can be obtained considering the pair $(l(q), B(q))$ which describes a
curve in a parametric form.
A similar procedure can be used to obtain the relation between any two variables with no need of simulation
tools. \\
\noindent 
% The proposed analysis technique is computationally efficient for this class of models such that it easily allows to investigate the influence of model parameters in the system behaviour. 
In \cite{Bookchapter} and \cite{MatBasGenCDC2004} it is shown that  the 
approximation error of the HB method for the analysis of this model is negligible when
compared
to results obtained by simulating the same model.

\section{Identification of the Tip-Sample Force Model \label{sect_identification}}

In this section we present methods to identify parameters of the hysteresis based
models developed in earlier.
As previously discussed, the frequency response $L(i\omega)$  is known since it can be independently estimated.
We also assume that the separation $l$ can be changed by means of the piezoactuator 
placed beneath the sample. 
Therefore, we can consider a set of $M$ experiments with different values
of $l$
\begin{equation}
	l_m:=l_0 + md \qquad m=1, ..., M
\end{equation}
where $l_0$ is a fixed offset and $d>0$  is a suitable separation step.
For every $l_m$ the quantities $A_m, B_m, \phi_m$ can be evaluated from the measured
signal $y(t)$ after it has reached its steady state, and $q_m$ can be computed from (\ref{qdefinition}).\\
The functions (\ref{generalclass}) chosen to model the interaction have the useful property
that it is linear in the parameters  $K_n^-$ and $K_n^+$, or in virtue of \ref{Sigma} and \ref{Delta},
$\Sigma_n$ and $\Delta_n$. The linear dependence on the parameters
aids their identification using the harmonic balance relations (\ref{autosostentamento2}).
The first order harmonic balance equations lead to a set of $M$ linear equations in the $2N$
unknown variables $\Sigma_n$ and $\Delta_n$
\begin{equation}\left\{
	\begin{array}{l}
%		A = \sum_{n=1}^{N} \Sigma_n \pi_n(q) \\
		\Gamma\cos(\phi_m)-\mathrm{Im}[L^{-1}(i\omega)]B_m=
				\sum_{n=1}^{N} \Sigma_n \frac{S_n(q_m)}{B^{n+1}} \\
		\Gamma\sin(\phi_m)-\mathrm{Re}[L^{-1}(i\omega)]B_m=
				\sum_{n=1}^{N} \Delta_n \frac{T_n(q_m)}{B^{n+1}} \\
	\end{array}\right.
		\qquad m= 1,... , M
\end{equation}
Assuming that there are $M>2N$ experimental measures and
adopting a more compact notation, we can write two independent matrix equations
\begin{equation}
	\begin{array}{l}\label{identequation}
		P_S{\Sigma} = Q_{S} \\
		P_D{\Delta} = Q_{D}
	\end{array}
\end{equation}
where
\begin{equation}
	\begin{array}{l}
	\Sigma:=\left( \begin{array}{l}
			\Sigma_1\\ \vdots\\  \Sigma_N
	         \end{array}
	 	\right)\qquad\qquad
	\Delta:=\left( \begin{array}{l}
			\Delta_1\\ \vdots\\  \Delta_N
	         \end{array}
	 	\right)
	\end{array}
\end{equation}
are the unknown vectors and 
\begin{equation}
	\begin{array}{cc}
   P_S[m,n]:=\frac{S_n(q_m)}{B_m^{n+1}}~~ & 
				Q_S[m]:=\Gamma\cos(\phi_m)-\mathrm{Re}[L^{-1}(i\omega)]B_m\\
   P_D[m,n]:=\frac{T_n(q_m)}{B_m^{n+1}}~~
				 &  Q_D[m]:=\Gamma\sin(\phi_m)-\mathrm{Im}[L^{-1}(i\omega)]B_m
	\end{array}
\end{equation}
are constant matrices.\\
Since the number of equations is greater than the number of unknowns, (\ref{identequation}) 
is not expected to be feasible.
A common strategy is to find the set of parameters
which better ``fits'' the equations according to the quadratic cost function
\begin{equation}
	V(l_0, \Sigma, \Delta)=\Vert Q_S - P_S \Sigma  \Vert^2 + \Vert Q_D - P_D \Delta  \Vert^2,
\end{equation}
where we have stressed the dependence on the offset $l_0$ since it is not apriori known.
Thus, the optimal values $\tilde \Sigma$ and $\tilde \Delta$ can be evaluated casting an optimization
problem which also takes into account the constraints (\ref{constraints})
\begin{equation}\label{OptProblem}
	\begin{array}{l}
		\displaystyle{
			(\tilde \Sigma(l_0), \tilde \Delta(l_0) )=\arg \min_{\Sigma, \Delta} V(l_0, \Sigma, \Delta)
		}\\
% 			~\\
			\qquad \mathrm{subject~to}\\
% 			~\\
			\Delta \leq 0.
		\end{array}
\end{equation}
We remark that forcing the condition $\Delta=0$ in (\ref{OptProblem}) is equivalent to the assumption
of a interaction force with no hysteresis and therefore conservative.\\
Given $l_0$, problem (\ref{OptProblem}) is a quadratic optimization problem with linear constraints.
Many algorithms are known in literature to determine its solution
$(\tilde \Sigma(l_0), \tilde \Delta(l_0))$ \cite{Boyd}.
Finally, we can estimate the offset $l_0$ by solution of the following problem
\begin{equation}
	\begin{array}{l}
		\displaystyle{
			\tilde l_0=\arg \min_{l_0} V(l_0, \tilde \Sigma(l_0), \tilde \Delta(l_0),)
		}\\
	\end{array}
\end{equation}
which is another minimization over a scalar variable, solvable using a grid strategy.
The identified parameters are $(\tilde \Sigma(\tilde l_0), \tilde \Delta(\tilde l_0))$.

\section{Experimental Results}\label{sect_expresult}
\begin{figure}[hbt]
\begin{tabular}{c}
%  \begin{center}
\psfrag{separation}[][l]{separation [nm]}
\psfrag{amplitude}[][l]{amplitude [nm]}
     \qquad\qquad \includegraphics[width=0.6\columnwidth]{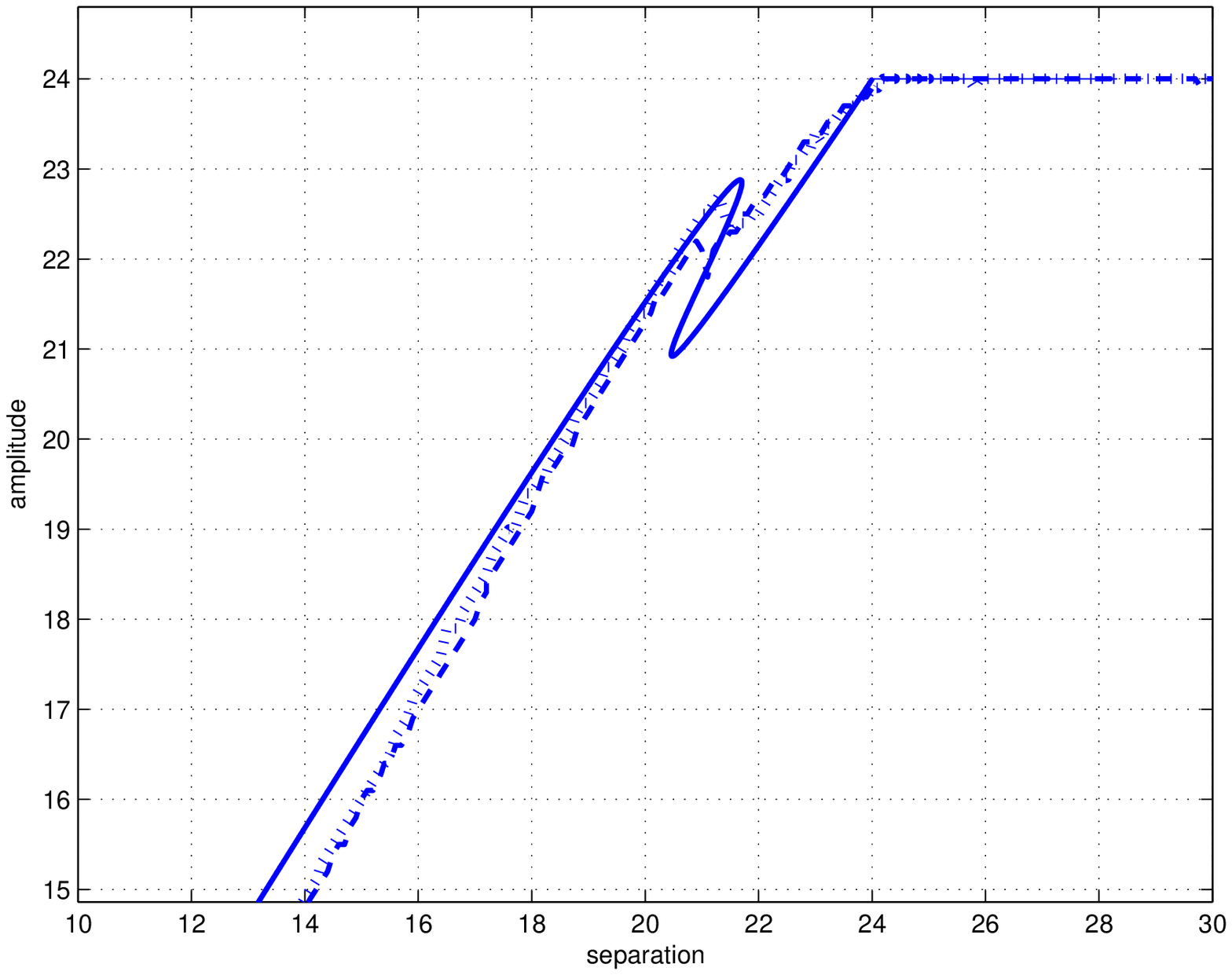}\\ 
	\qquad\qquad (a)\\
%\end{center}
	\\
\psfrag{separation}[][l]{separation [nm]}
\psfrag{phase}[][l]{phase [rad]}
     \qquad\qquad \includegraphics[width=0.6\columnwidth]{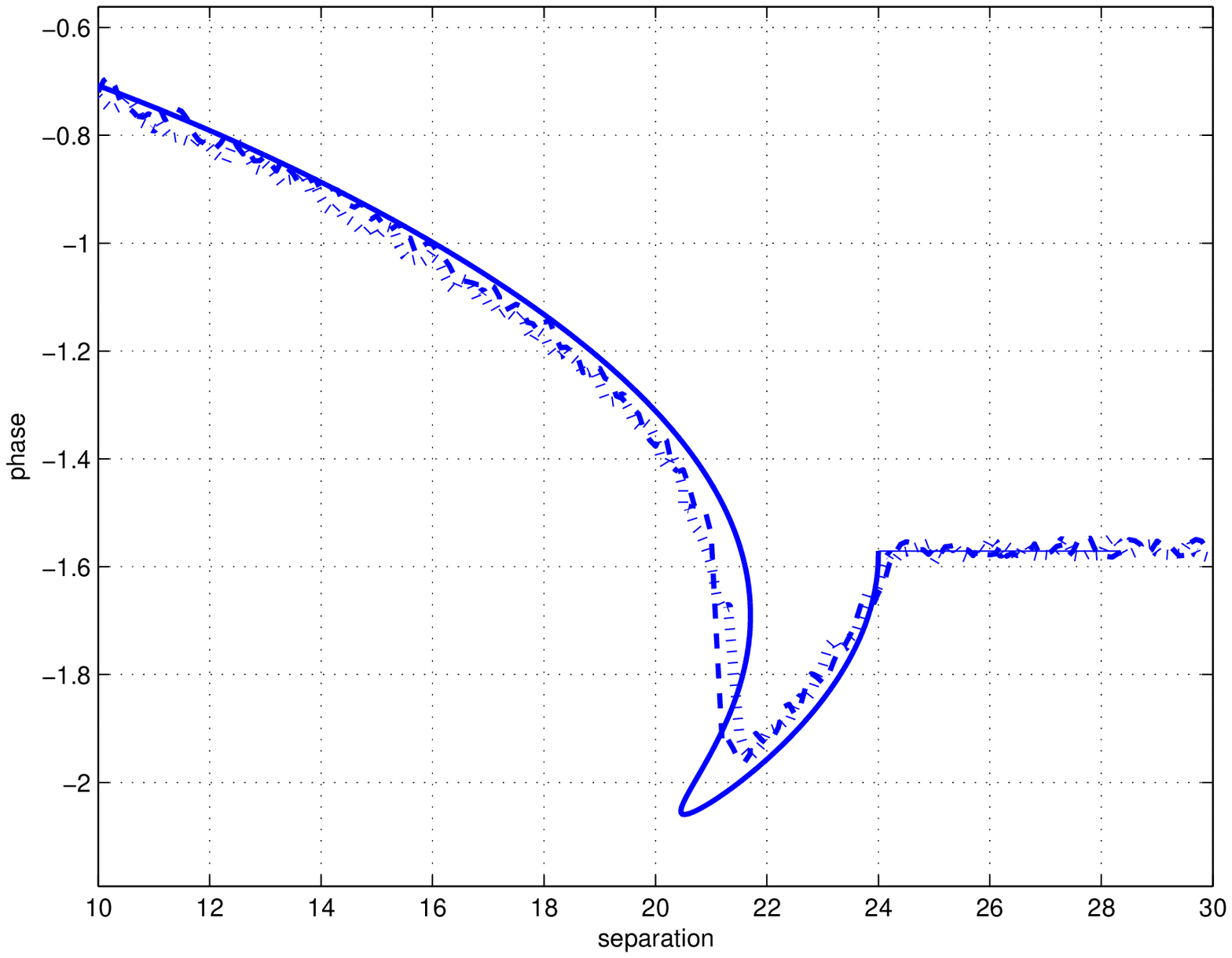}\\
	\qquad\qquad (b)\\
\end{tabular}
     \caption{Experimental separation-amplitude (a) and separation-phase (b) curves fitted
		 using the piecewise linear model for the interaction force.
		Solid line is the curve obtained by the model; the dashed and the dotted ones are
			 the experimental approach and retract curves respectively.}
			\label{fig_PW}
\end{figure}

\begin{figure}[hbt]
\begin{tabular}{c}
%  \begin{center}
\psfrag{separation}[][l]{separation [nm]}
\psfrag{amplitude}[][l]{amplitude [nm]}
     \qquad\qquad \includegraphics[width=0.6\columnwidth]{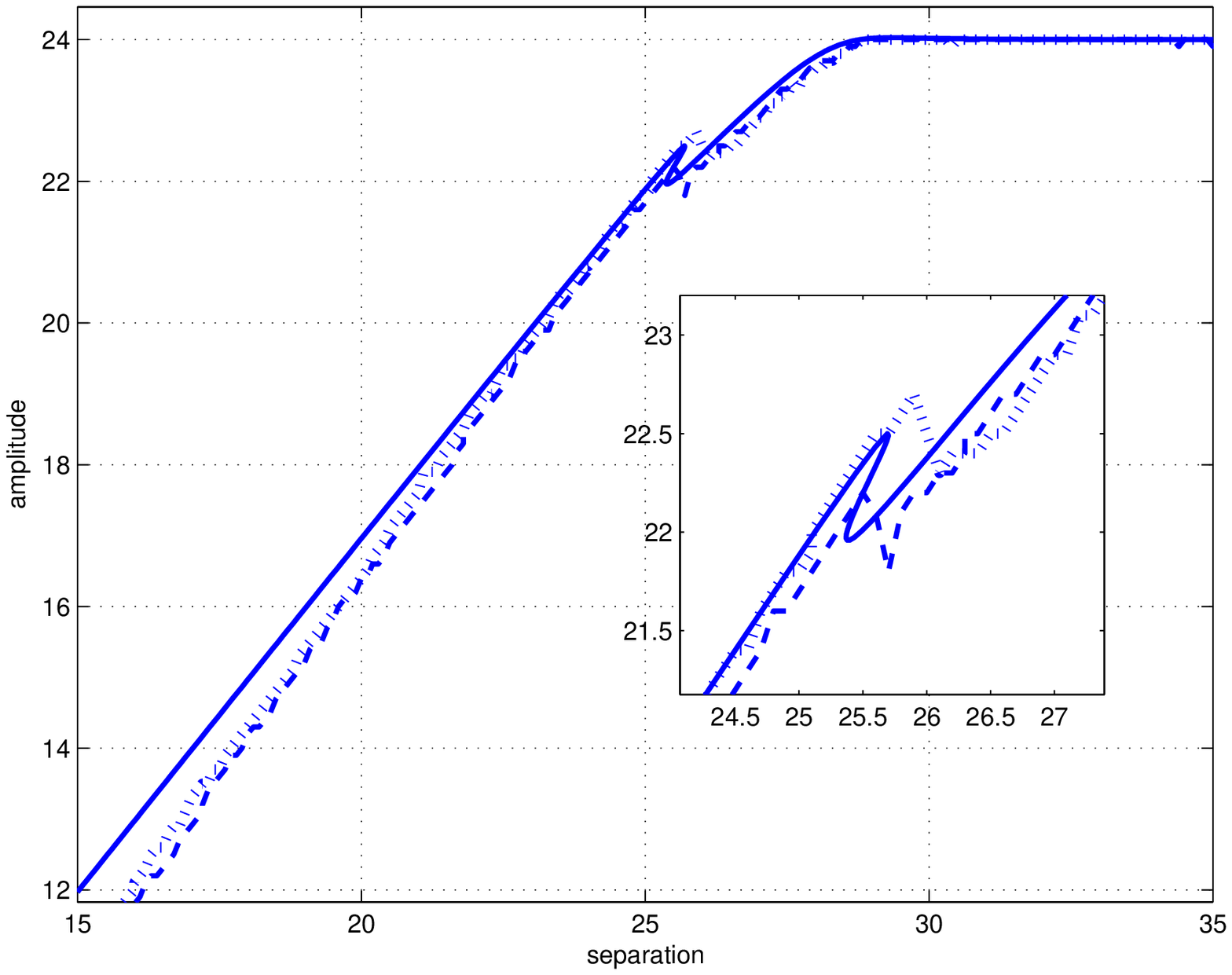}\\ 
	\qquad\qquad (a)\\
%\end{center}
	\\
\psfrag{separation}[][l]{separation [nm]}
\psfrag{phase}[][l]{phase [rad]}
     \qquad\qquad \includegraphics[width=0.6\columnwidth]{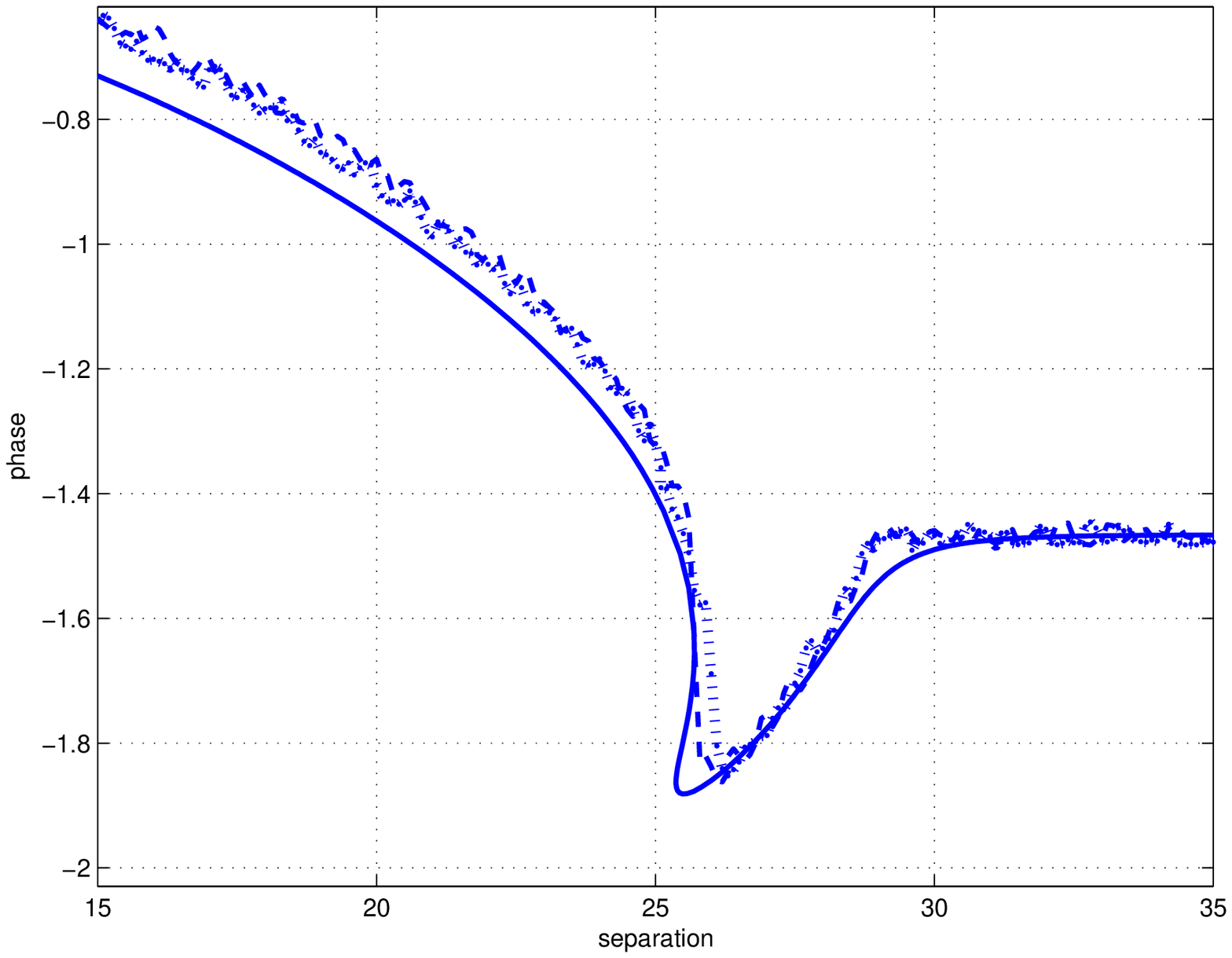}\\
	\qquad\qquad (b)\\
\end{tabular}
     \caption{Experimental separation-amplitude (a) and separation-phase (b) curves fitted using the
			Lennard-Jones model for the interaction force.
			Solid line is the curve obtained by the model; the dashed and the dotted ones are
			 the experimental approach and retract curves respectively.}\label{sepamp713}
\end{figure}

An atomic force microscope was operated in dynamic mode using a silicon cantilever of 225 
$\mu$m in length.
Using a thermal-response based approach
the cantilever has been identified with a second order linear oscillator
with natural frequency  $\omega_n=2\pi  73.881 ~rad/s$ and damping  factor $\xi=0.00378$.
The spring constant $K$ of the cantilever has been estimated to be $4 nN/nm$.
% In Figure \ref{ThResponse} the result of an identification  based on the thermal response is shown.
% \begin{figure}[hbt]
%    \begin{center}
%      \includegraphics[width=0.65\columnwidth]{ThResponse}\\
%      \caption{The system thermal response (solid line) is very well fitted by a
%         fourth order model (dashed line)\label{ThResponse}}
%     \end{center}
% \end{figure}
A sinusoidal voltage with frequency $\omega_n$ has been applied to the dither piezo in order to
make the cantilever oscillate.
Experiments were performed on a silicon wafer sample.
The separation amplitude curve has been experimentally measured during the two phases:
approach and retract. Both curves show a jump phenomenon occurring at two close 
but different values of the separation (dashed and dotted curves in Figure \ref{fig_PW}). 
Such phenomena are present and documented in literature \cite{Attrattive}.
The identification technique described in the previous section has been employed using the data
obtained during the retract phase only, while the data acquired during the approaching phase
have been used for validation purposes.
The results obtained using the piecewise linear interaction model are reported in Figure \ref{fig_PW}
(solid curve).
% The retract curve is well fitted by the model, as it should be expected since those data have been used in the
% identification. 
The retract curve is well explained by the model data; this is not surprising as the data
used to obtain the model parameters is the retract phase data.
For the approach curve, a jump phenomenon occuring at a different separation is qualitatively well-predicted, 
but it can be argued that it is not quantitatively satisfactory.\\
As a second case, the following simplified model than (\ref{generalclass}) is used
\begin{equation}\label{LJpotential} 
	h(\delta,\dot \delta) =
		\left\{\begin{array}{l}
			\frac{K_{7}^-}{\delta^{7}}+\frac{K_{13}^-}{\delta^{13}}
				\qquad \mathrm{if}~\dot \delta < 0 \\
			\frac{K_{7}^+}{\delta^{7}}+\frac{K_{13}^+}{\delta^{13}}
				\qquad \mathrm{if}~\dot \delta > 0 \\
		       \end{array}
		\right.
% 		\qquad\mathrm{with}~~n_r>n_a
\end{equation}
that is a standard $6-12$ Lennard-Jones potential function with a hysteresis dissipation.
Identification results are shown in Figure \ref{sepamp713}.
As it is evident from the figure, the model predicts the discontinuity in the approach phase
of the force curve accurately. The phase of the first harmonic is also predicted well by the
model.\\
% As it is shown, the retracting curve is again well-fitted by the identified one.
% However, what actually confirms  the better description provided by this model is the fact that 
% the jump point in the approaching curve is now predicted with a quite good accuracy. 
% The phase diagram equally presents a satisfactory agreement.\\
As remarked in Figures  \ref{fig_PW} and  \ref{sepamp713}, 
harmonic balance has also allowed to reveal the presence of instable periodic orbits 
(in the region in between the two jump points) and to clearly explain bifurcation phenomena in the system.

\begin{figure}
	\begin{center}
	\psfrag{force}{force [normalized]}
	\psfrag{separation}{separation [nm]}
	\includegraphics[width=0.6\columnwidth]{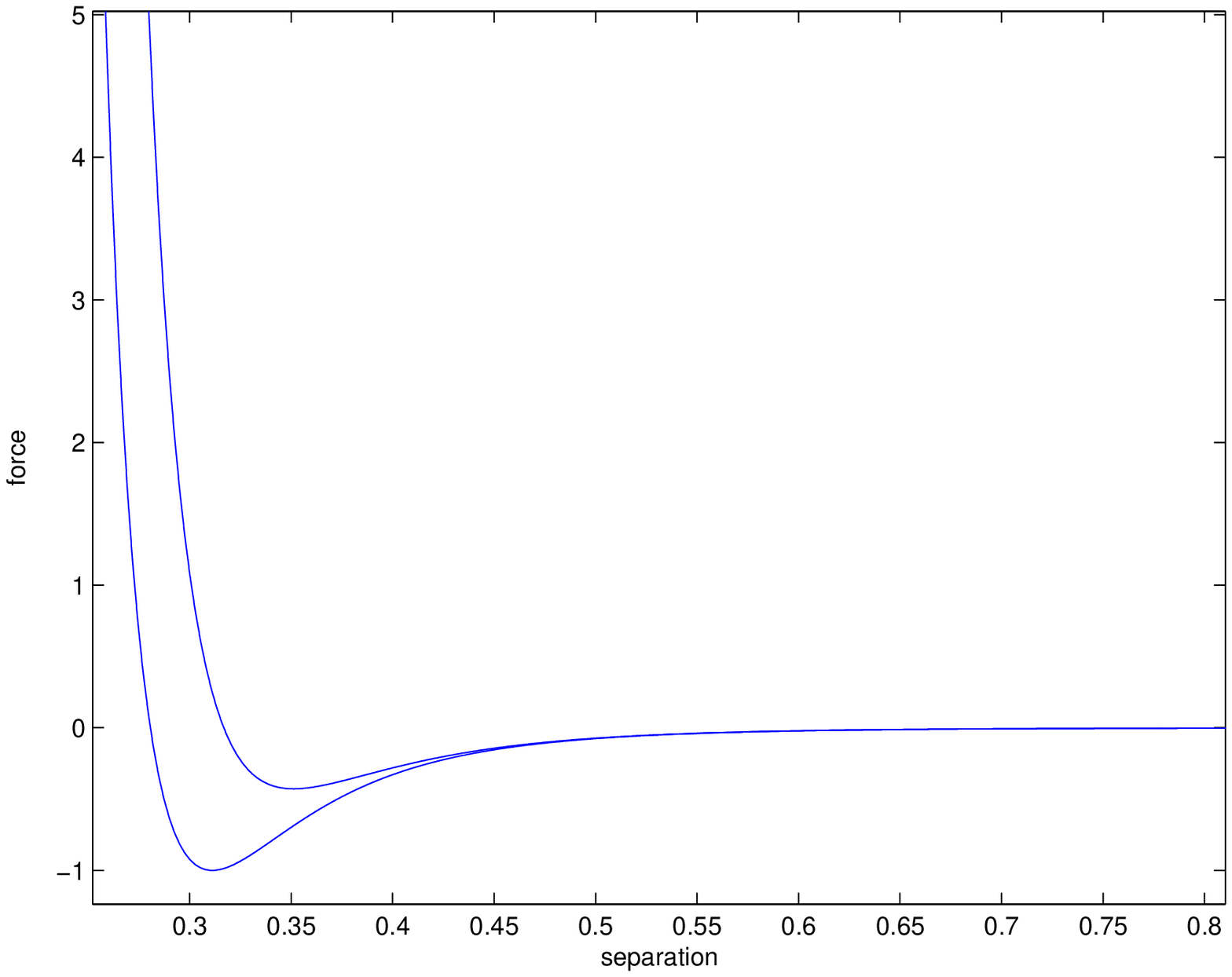}
	\caption{Identified dynamic Lennard-Jones force curve
			with hysteresis. \label{LJidentcurve}}
	\end{center}
\end{figure}

\section{Conclusions \label{Comments}}
In the paper we have proposed a class of models for tip-sample interaction in atomic force microscopy via impact dynamics. 
The use of a hysteresis can be well combined with harmonic balance techniques
for the analysis of oscillatory behaviour to provide interesting insights into the dynamics.
For instance, the presence of jump phenomena
discovered in many experiments is well-predicted and explained.
The suggested method is based on a first order harmonic approximation and gives good quantitative results since the linear part of the considered Lur'e system shows a sharp filtering effect near the resonance
frequency.
In such a situation, the Harmonic Balance technique has advantages over standard numerical approaches since it requires a computational effort much smaller than the one required by simulation tools.

\bibliographystyle{unsrt}
\bibliography{biblio}

\end{document}